\documentclass{aa} 
\usepackage{graphicx,natbib} 
\usepackage{txfonts} 
\bibpunct{(}{)}{;}{a}{}{,} 
 
\def\h2o{H$_2$O}
\def\source{OH231.8+4.2}
 
\begin{document} 
 
\title{Rotten Egg Nebula: The magnetic field of a binary evolved star} 
 
\author{Leal-Ferreira, M. L.\inst{1} \and Vlemmings, W. H. T.\inst{1,2} 
\and Diamond, P. J.\inst{3} \and Kemball, A.\inst{4,5} \and Amiri, N.\inst{6,7} 
\and Desmurs, J. -F.\inst{8}}

\institute{Argelander Institute f\"ur Astronomie, Universit\"at Bonn, 
Auf dem H\"ugel 71, 53121 Bonn, Germany\\ 
\email{ferreira@astro.uni-bonn.de, wouter@astro.uni-bonn.de} 
\and Department of Earth and Space Sciences, Chalmers University of Technology, Onsala Space Observatory, SE-439 92 Onsala, Sweden
\and JBCA, School of Physics \& Astronomy, Manchester University, M13 9PL, 
UK 
\and Department of Astronomy, University of Illinois at Urbana-Champaign, 
1002 West Green Street, Urbana, IL 61801, USA 
\and National Center for Supercomputing Applications, University of 
Illinois at Urbana-Champaign, 605 East Springfield Avenue, Champaign, IL 
61820, USA 
\and Joint Institute for VLBI in Europe, PO Box 2, 7990 AA Dwingeloo, 
The Netherlands 
\and Leiden Observatory, Leiden University, PO Box 9513, 2300 RA Leiden, 
The Netherlands 
\and Observatorio Astron\'omico Nacional, C/Alfonso XII 3, 28014 Madrid, 
Spain}
\date{received , accepted} 
\authorrunning{Leal-Ferreira et al.} 
\titlerunning{The magnetic field of the pPN OH231.8} 
 
\abstract 
% context heading (optional) 
{Most of the Planetary Nebulae (PNe) observed are not spherical. The loss 
of spherical symmetry occurs somewhere between the Asymptotic Giant Branch 
(AGB) phase and the PNe phase. The cause of this change of morphology is not 
yet well understood, but magnetic fields are one of the possible agents. The origin 
of the magnetic field remains to be determined, and potentially requires the 
presence of a massive companion to the AGB star. {Therefore, further 
detections of the magnetic field around evolved stars, and in particular those
thought to be part of a binary system, are crucial to improve our 
understanding of the origin and role of magnetism during the late
stages of stellar evolution.  One such binary is the pre-PN OH231.8+4.2, 
around which a magnetic field has previously been detected in the OH maser region of the 
outer circumstellar envelope.}} 
% aims heading (mandatory) 
{We aim to detect and infer the properties of the magnetic field of the 
pre-PN OH231.8+4.2 in the \h2o maser region 
that probes the region close to the central star. This source is a confirmed binary with collimated outflows and an 
envelope containing several maser species.} 
% methods heading (mandatory) 
{In this work we observed the 
6$_{1,6}-$5$_{2,3}$ \h2o maser rotational transition to determine its 
linear and circular polarization. As a result of Zeeman splitting, the properties of the magnetic field can be 
derived from maser polarization analysis. The \h2o maser emissions of \source 
~are located within the inner regions of the source (at a few tens of AU).} 
% results heading (mandatory) 
{We detected 30 \h2o maser features around \source. The masers occur in two 
distinct regions that are moving apart with a velocity on the sky of 
2.3~mas/year. Taking into account the inclination angle of the source with the 
line of sight, this corresponds to an average separation velocity of 21~km/s. 
Based on the velocity gradient of the maser emission, the masers 
appear to be dragged along the direction of the nebula jet. Linear 
polarization is present in three of the features, and circular polarization is 
detected in the two brightest features. The circular polarization results imply a magnetic field strength of  $|$B$_{||}|$$\sim$45~mG.} 
%conclusions heading 
{We confirm the presence of a magnetic field around \source, and 
report the first measurements of its strength within a few tens of AU of the 
stellar pair. Assuming a toroidal magnetic field, this imples  
B$\sim$2.5~G on the stellar surface. The morphology of the 
field is not yet determined, but the high scatter found in the directions 
of the linear polarization vectors could indicate that the masers occur near 
the tangent points of a toroidal field.} 
 
 
\maketitle 
 
\section{Introduction} 

During the final stages of their evolution, low and intermediate 
mass stars (0.8$-$8~M$_\odot$) evolve from the Asymptotic Giant Branch 
(AGB) phase to Planetary Nebulae (PNe). During this transition most of 
these objects lose their spherical symmetry. The process responsible 
for the change of morphology is, so far, not well understood. 
\cite{bujarrabal01} have shown that, in most cases, the radiation 
pressure does not have enough energy to drive the acceleration of the 
fast bipolar flows observed in pre-PNe (pPNe). Moreover, the mass 
ejection on a preferred axis requires an extra agent to collimate 
the flow. The possible mechanisms that could both provide the missing energy 
to drive the fast flows and that could collimate it on a preferred axis 
are: (i)~a companion to the star (either a binary stellar companion or 
massive planet) and its tidal forces, (ii)~disk interaction and 
(iii)~magnetic fields - or a combination of these \citep[][~and 
references therein]{balick02,frank07,nordhaus07}.

From magneto-hydrodynamic (MHD) simulations, \cite{garcia99} concluded 
that magnetic fields can indeed have a pronounced effect on shaping stars 
beyond the AGB stage and that, working together with rotation, they can 
produce collimated bipolar nebulae and jets. Since then, several other 
MHD simulations have been designed to investigate how magnetic fields can 
shape a pPN and, consequently, a Planetary Nebula \citep[e.g.,][]{garcia05,
garcia08,dennis09}. 

Observations of the magnetic fields around evolved stars, however, are 
still rare. 
Most current magnetic field measurements are performed using Zeeman splitting 
\citep{zeeman97} of maser lines \citep[e.g.,][]{vlemmings01,vlemmings06}.
Each maser species requires different physical conditions to occur. 
Therefore, different masers originate in different locations of the 
studied object. The SiO masers are found close to the central star (CS), in 
regions with a temperature of $\sim$1300~K. The OH masers are generally 
expected to occur further out, at a few hundred AU from the CS. 
For AGB stars, the \h2o maser emission is expected to be intermediate between 
these two regions; but for a pPN/PN it can be found at a similar or greater 
radius than the OH maser emission \citep[][]{cohen87,reid97,vlemmings06b}.

In this work, we aim to investigate the polarization of the \h2o maser 
emission toward the pPN OH231.8+4.2 (Rotten Egg Nebula; Calabash Nebula), 
and infer the properties of its magnetic field. Located at a distance of 
$\sim$1540~pc \citep[][ updated from private communication]{choi11}, the bipolar nebula \source ~contains a 
binary system in its core, where a Main Sequence type~A star accompanies 
an evolved star - the Mira variable QX~Pup \citep[][]{sanchez04}. The SiO, 
\h2o and OH maser emission toward the Rotten Egg Nebula have been observed 
several times before \citep[e.g., ][]{morris82,gomez01,sanchez02,desmurs07,etoka08,etoka09}. 
These prior works have shown that the spatial distribution of both the 
SiO and OH maser emission lie perpendicular to the nebular symmetry 
axis, consistent with the presence of a torus around the equator of the central 
star. However, the \h2o masers (located at only $\sim$40~AU from the CS) do 
not follow the same pattern, and are instead located toward the bipolar 
structure. Few of the prior studies included polarization measurements. 
\cite{sanchez02} note that their SiO polarization maps are not reliable and 
\cite{gomez01} found no circular polarization in the OH maser emission. 
\cite{etoka08,etoka09}, however, found both circular and linear 
polarization in the OH masers, and conclude that a well-organized magnetic 
field seems to be flaring out in the same direction as the outflow.

This paper is structured as follows: Section~2 describes the 
observations, data reduction, and calibration,  and Section~3  includes a presentation of the 
results. The results are discussed in Section~4 and the concluding analysis is presented in Section 5. 

\section{Observations and Data Reduction} 
 
The observations of the pPN \source ~were carried out on March 1$^{st}$, 
2009, using the NRAO\footnote{The National Radio Astronomy Observatory 
(NRAO) is a facility of the National Science Foundation operated under 
cooperative agreement by Associated Universities, Inc.} Very Long Baseline 
Array (VLBA), under project code BV067A. We observed the \h2o 6$_{1,6}-$5$_{2,3}$ 
rotational transition, at an adopted rest frequency of 22.235081~GHz. We used 2 
baseband filters of 1~MHz bandwidth and, following the VLBA \h2o observations 
of \cite{desmurs07} (hereafter; D07), the filters were centered at V$_{lsr}$ 
44.0~km~s$^{-1}$ and 26.0~km~s$^{-1}$ respectively. This was done in order to detect both 
the emission from the northern (NReg; $\sim$30~mas from the CS) and 
southern regions (SReg; $\sim$40~mas from the CS) of the source. The 
observations were centered on 
RA$_0$(J2000)~=~07$^h$42$^m$16.93 and Dec$_0$(J2000)~=~$-$14$^\circ$42'50''.2. 
Both a lower and a higher spectral resolution correlation were performed. 
The former was undertaken in full polarization mode (to obtain all 4 Stokes 
correlations: $RR$, $LL$, $RL$ and $LR$) over 128 channels, with a 
resulting nominal channel width of 0.104~km/s. The latter correlation was performed in dual polarization 
mode (to obtain $RR$ and $LL$) over 512 channels, with a resulting nominal channel width of 
0.026~km/s. The synthesized beam size was $\sim$1.7$\times$0.9~mas. The 
observations spanned 8 hours, of which 4.7 hours were spent on \source. 

For calibration purposes, we observed J0854+2006. This reference source was 
used to perform bandpass, phase and polarization calibration. We used the 
Astronomical Image Processing Software Package (AIPS) and followed the data 
reduction procedure documented by \cite{kemball95},  and which has 
been successfully adopted by several authors \citep[e.g., ][]{vlemmings01,surcis11}.

Using the low resolution data, we created image cubes for the Stokes 
parameters $I$, $Q$, $U$ and $V$. The noise level measured from these 
image cubes ranges between $\sim$3 and $\sim$5~mJy/beam in the 
emission free channels. From the Stokes $Q$ and $U$ images, we derived image 
cubes for the linear polarization intensity (P) and the linear polarization 
direction (also referred to as electric vector position angle, EVPA). 

Using the high resolution data, we created image cubes for the Stokes
parameters $I$ and $V$. The noise level measured from these image
cubes lay between $\sim$6 and $\sim$10~mJy/beam in the emission free
channels.  To estimate the polarization leakage and consequently
  the minimum believable fractional linear polarization we determined
  the Stokes $Q$ and $U$ limits for the brightest maser feature with
  no detected polarization. We conclude that, at $3\sigma$, the minimum detected
  fractional polarization is $\sim$0.1$\%$, a value that is consistent
  with the self-calibration method used during the data reduction, 
  described in further detail by \citet{leppanen95}.

In our analysis we adopted a signal-to-noise ratio cutoff of four 
times the measured noise (4$\sigma$). We define a maser feature to be 
successfully detected when maser spots located at similar spatial positions 
(within the beam size) appear and survive the threshold cut in at least 
3 consecutive channels. The position of the maser feature is then defined 
by the emission in the channel where the peak appears \citep[see e.g.][]{richards11}.

\section{Results} 
\subsection{Masers Location} 
 
\begin{figure*}
\centering
\begin{tabular}{cc}
  \includegraphics[width = 75.5 mm]{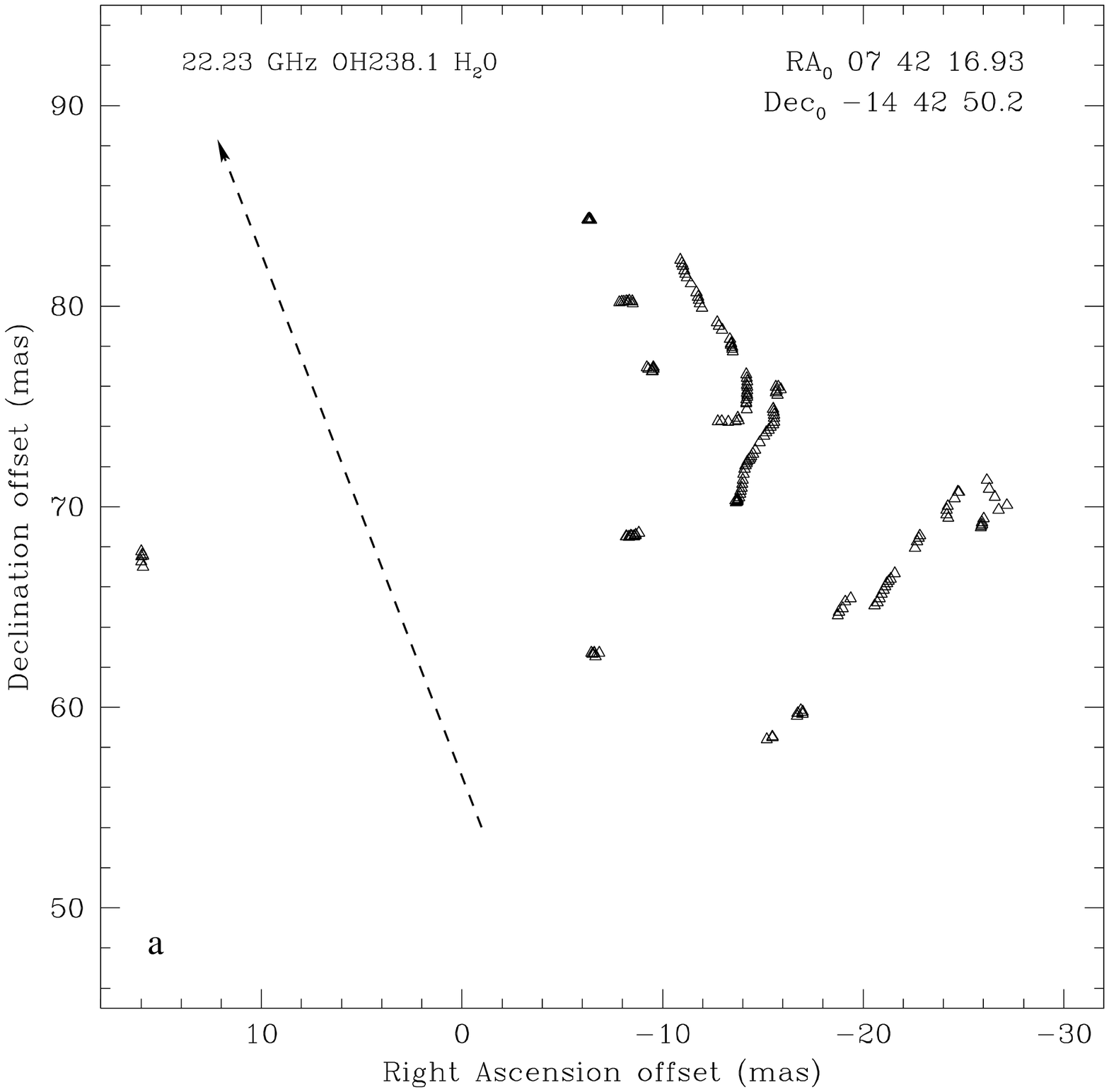} &
  \includegraphics[width = 94 mm]{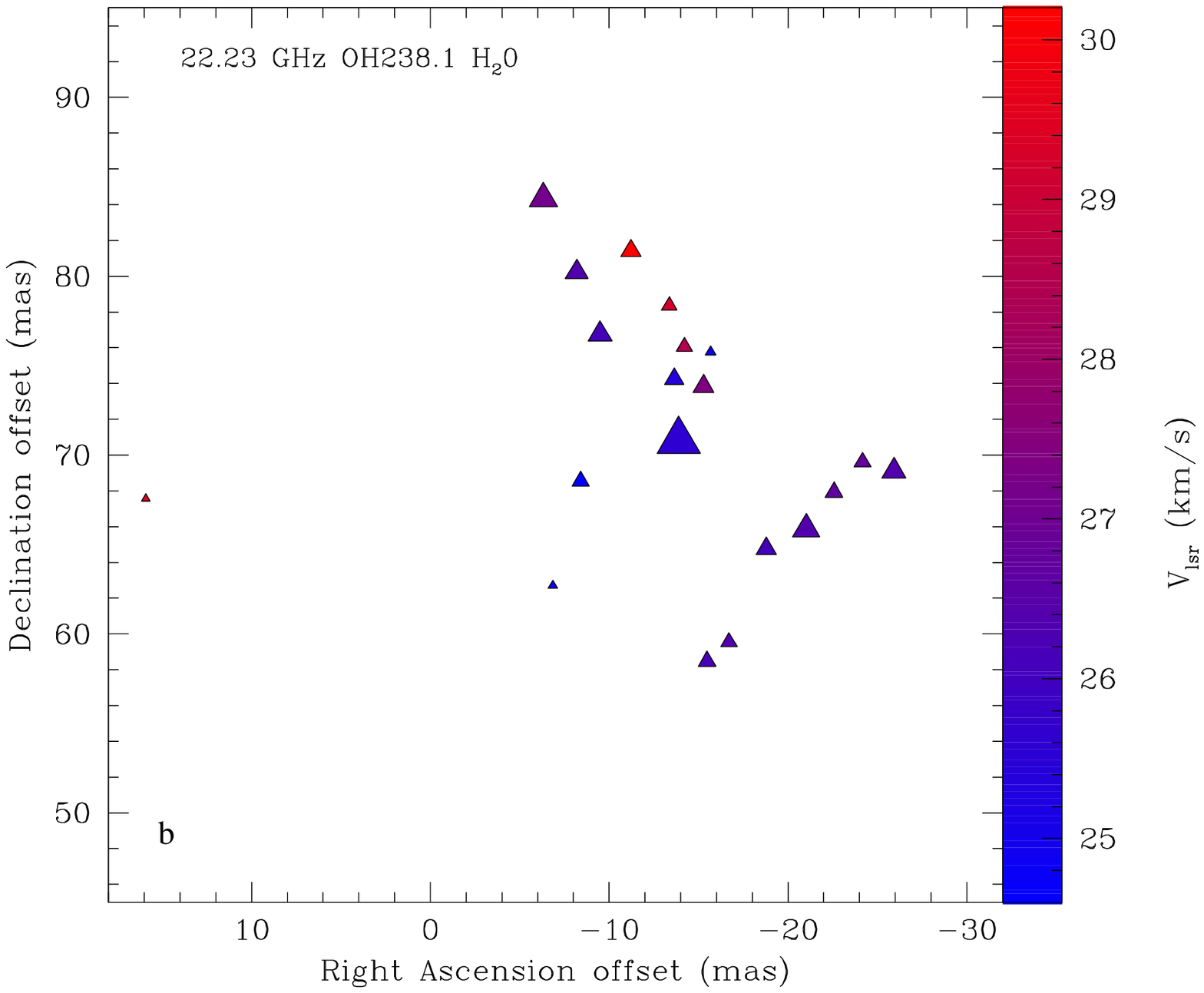} \\
\\
  \includegraphics[width = 79 mm]{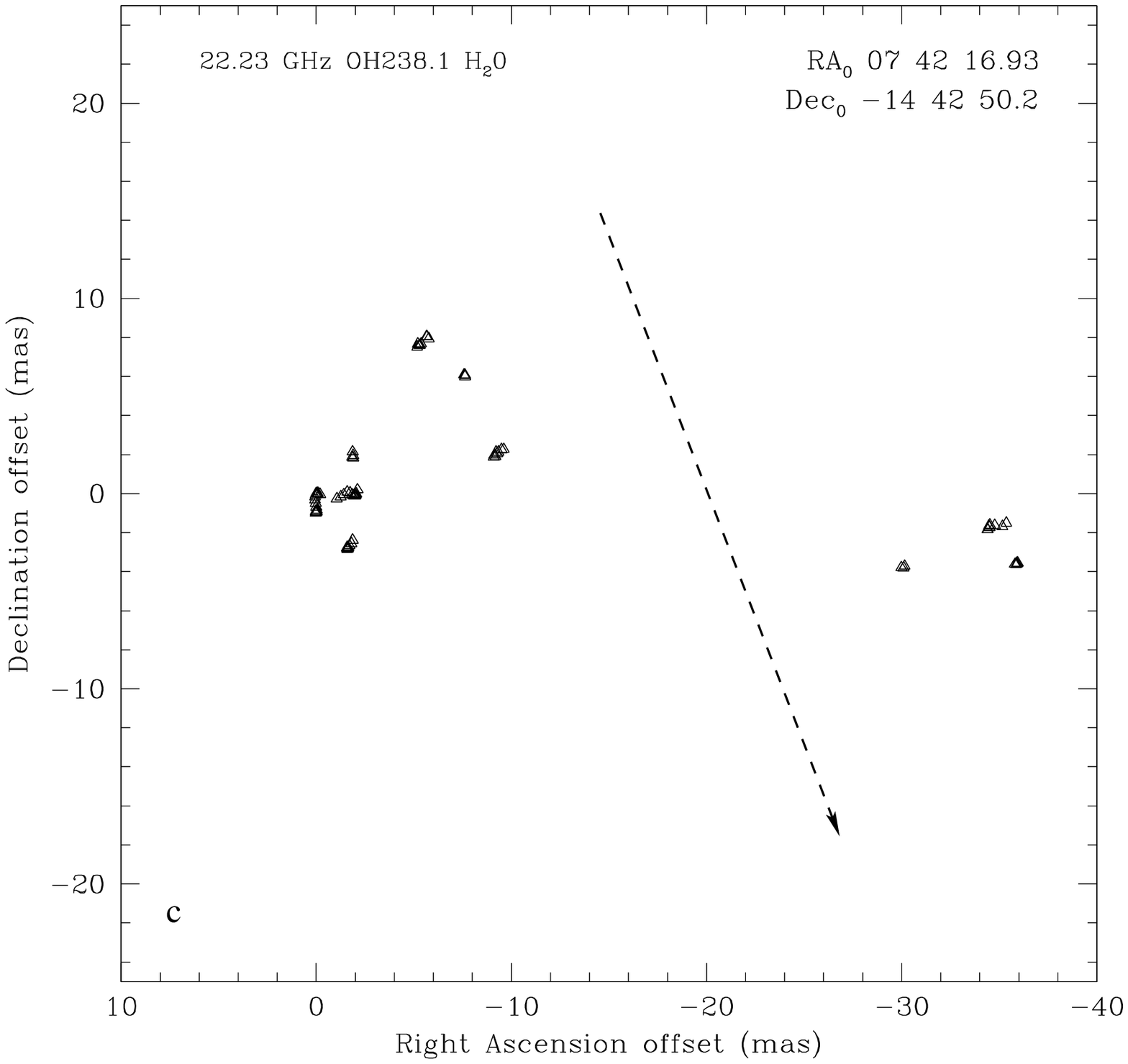} & 
  \includegraphics[width = 94 mm]{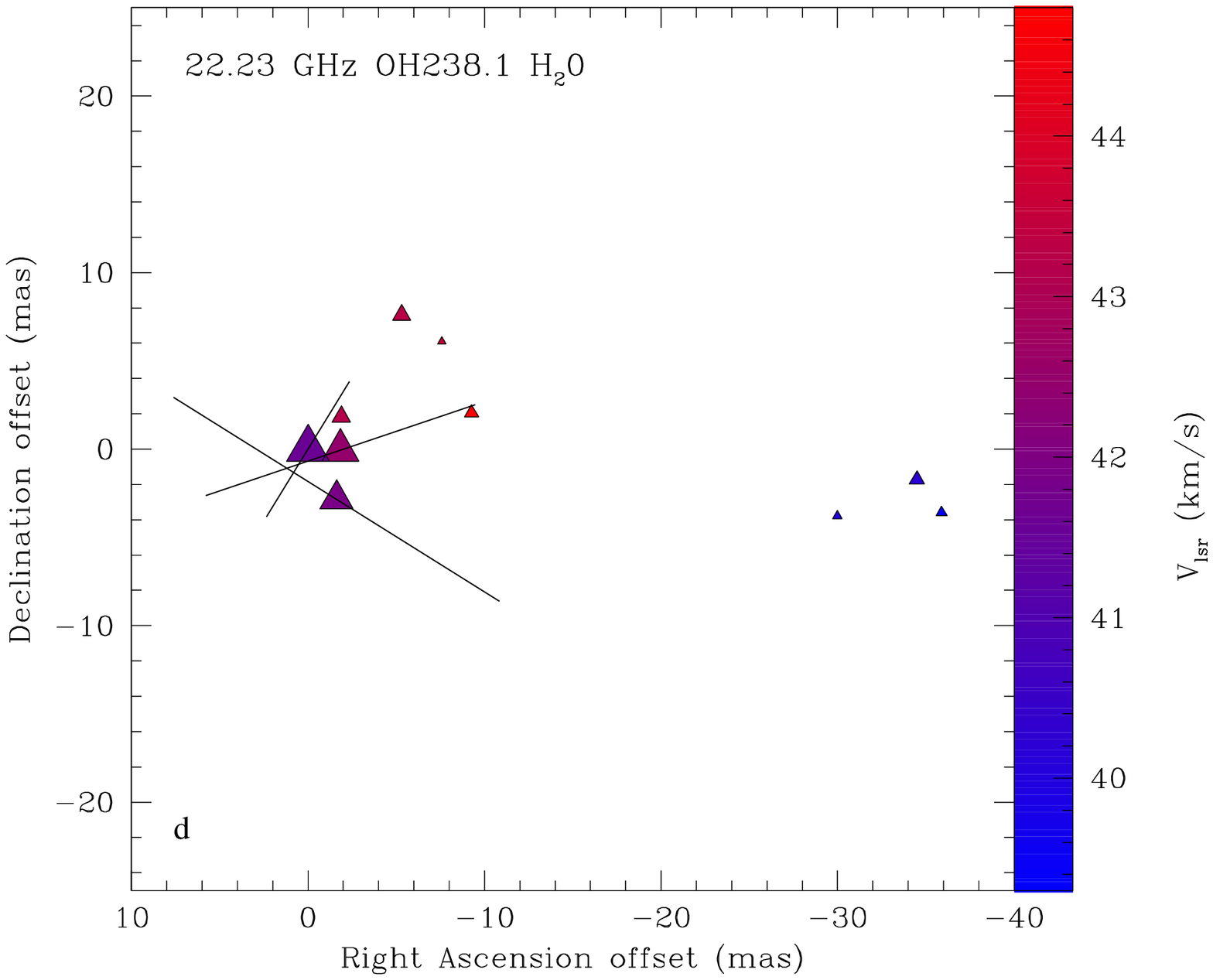} 
\end{tabular}
\caption[]{Plots a and b (top) corresponds to NReg, and plots c and d (bottom) 
to SReg. Plots a and c (left) show the position of all the measured maser 
spots, and plots b and d (right) present the maser features. On plots a and 
c the dashed vectors indicate the direction of the pPN jet. On plots b and 
d the size of the triangles are scaled by the maser fluxes, and the color 
scale is related to the velocity. The black lines represent the EVPA 
and intensity of the linear polarization.}
  \label{maserpos}
\end{figure*}

We detected 30 maser features in total, 20 of them located in the NReg 
and the remaining 10 located in the SReg. In Figure \ref{maserpos} we 
present four plots: 1.a and 1.b correspond to the NReg, and 1.c and 1.d 
to the SReg. Plots (a) and (c) show the distribution of all 252 maser 
spots that comprise the 30 maser features. The dashed vectors indicate 
the direction of the pPN jet. Plots (b) and (d) show the maser features, 
scaled by their flux densities (proportional to the size of the triangles), and color-coded 
according to their line-of-sight velocities. The black line segments, scaled 
in size by the degree of polarization, show the EVPA (see section 
\ref{linpol}). The origin (0,0) of all four plots is centered on the brightest 
feature we detected, located in the SReg.

\begin{figure}
\centering
\begin{tabular}{cc}
  \includegraphics[width = 87 mm]{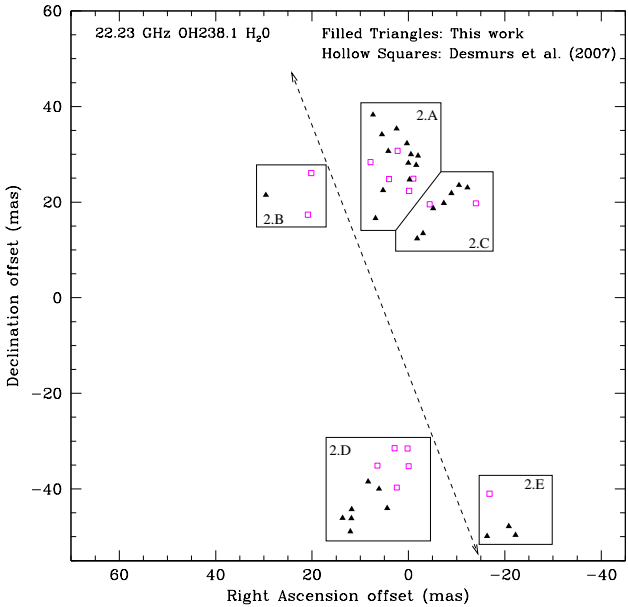} \\
\end{tabular}
\caption[]{Overplot of the maser features observed by D07 and the features 
presented in the present work. The filled triangles show the 30 features 
we observed and the hollow squares represent the observations from D07. The 
dashed vector indicate the direction of the pPN jet.}
  \label{masercomp}
\end{figure}

In order to aid the comparison of the observations presented by D07 with the current work, we over-plotted both observations in Figure~\ref{masercomp}. Since we 
did not use phase-referencing during our observations, unlike D07, we 
do not have a precise value for the absolute maser positions (but we do have 
accurate relative astrometric information). Therefore, to be able to make 
this comparison, we assumed a common origin point at the center of the maser 
emission. We determined this center point by taking the mean position 
over all the 30 maser features detected here, and over 15 of the 16 features 
detected by D07. We chose not to include maser feature number 16 from D07 
in the analysis, because it is clearly a spatial outlier in the field 
that contains the features we observed. Figure~\ref{masercomp} shows the 
entire field of view (NReg and SReg). The filled triangles represent the 
30 features detected in the current work and the hollow squares show the observations 
from D07. Five regions (2.A, 2.B, 2.C, 2.D and 2.E) are drawn on the figure
to facilitate their identification in the text.

In the NReg, the positions of the brightest feature and its surrounding features, are similar to those found by D07, except that the current observations detected more widespread features (region~2.A). On the eastern 
side of the brightest northern feature, at a separation of $\sim$30~mas, 
we detected only one weak feature, while two features were detected by D07 in this region, 
at $\sim$20~mas from the brightest maser (region~2.B). On the western
side of the main feature we detected a group of 7 features that together
seem to compose a single extended structure (region~2.C). This group of 
features does not appear in the observations of D07. Instead, they found 
only two features in that same region. Moreover, the emissions we detected in 
the NReg are contained within a much wider North-South area ($\sim$26~mas) 
than that in the D07 image ($\sim$13~mas).

In the SReg, the distribution of the features we observed is similar to 
those presented by D07, but appears to have moved away from the origin. 
Furthermore, we detected 10 features in this region, while D07 
detected only six features (considering only regions 2.D and 2.E). The detections in the current observations are 
spread over a wider East-West area ($\sim$36~mas) than\ that reported by D07 ($\sim$17~mas).

The 30 maser features and their properties are listed in Table~\ref{maserprop}. %From the first to the last column, respectively, the following maser parameters are presented: region of location (Reg); feature label (Label); right ascension offset, in reference to Figure~\ref{maserpos} ($\Delta\alpha$); declination offset, in reference to Figure~\ref{maserpos} ($\Delta\delta$); peak intensity ($I$); integrated intensity ($\int I$); radial velocity (V$_{lsr}$); P; EVPA.

\begin{table*}
  \centering
  \begin{minipage}{137mm}
    \caption{Maser features detected on \source}
    \begin{tabular}{@{}ccccccccc@{}}
      \hline
      Reg & Label &$\Delta\alpha$ (mas)&$\Delta\delta$ (mas)& $I$ (Jy/beam) & $\int$ $I$ (Jy) & V$_{lsr}$ (km/s) & P ($\%$) & EVPA ($^\circ$) \\
      \hline
      NReg & N.01 & $-$11.2057 & 81.4210 &  0.49 & 0.81 &   30.2   &  --             &   --              \\
      NReg & N.02 & +15.8905 & 67.5530 &  0.08 & 0.12 &   29.4   &  --             &   --              \\
      NReg & N.03 & $-$13.3862 & 78.3560 &  0.24 & 0.70 &   29.2   &  --             &   --              \\
      NReg & N.04 & $-$14.2320 & 76.0570 &  0.27 & 0.32 &   28.5   &  --             &   --              \\
      NReg & N.05 & $-$15.2988 & 73.8400 &  0.54 & 1.08 &   27.4   &  --             &   --              \\
      NReg & N.06 & $-$6.31373 & 84.3500 &  1.67 & 1.61 &   27.1   &  --             &   --              \\
      NReg & N.07 & $-$24.1890 & 69.6120 &  0.30 & 0.46 &   26.8   &  --             &   --              \\
      NReg & N.08 & $-$22.6036 & 67.9370 &  0.33 & 1.33 &   26.7   &  --             &   --              \\
      NReg & N.09 & $-$16.7119 & 59.5600 &  0.26 & 0.38 &   26.4   &  --             &   --              \\
      NReg & N.10 & $-$25.9197 & 69.1120 &  0.92 & 1.72 &   26.4   &  --             &   --              \\
      NReg & N.11 & $-$21.0261 & 65.8640 &  1.42 & 2.33 &   26.4   &  --             &   --              \\
      NReg & N.12 & $-$8.18628 & 80.2300 &  0.72 & 0.98 &   26.4   &  --             &   --              \\
      NReg & N.13 & $-$15.4815 & 58.4830 &  0.34 & 0.60 &   26.2   &  --             &   --              \\
      NReg & N.14 & $-$9.49584 & 76.7490 &  0.86 & 1.34 &   26.1   &  --             &   --              \\
      NReg & N.15 & $-$18.7994 & 64.7640 &  0.49 & 0.79 &   26.1   &  --             &   --              \\
      NReg & N.16 & $-$13.9092 & 70.7840 & 16.47 &19.24 &   25.6   &  --             &   --              \\
      NReg & N.17 & $-$13.6490 & 74.2530 &  0.43 & 0.73 &   25.4   &  --             &   --              \\
      NReg & N.18 & $-$6.86105 & 62.7080 &  0.09 & 0.11 &   24.9   &  --             &   --              \\
      NReg & N.19 & $-$15.6860 & 75.7780 &  0.11 & 0.13 &   24.8   &  --             &   --              \\
      NReg & N.20 & $-$8.41686 & 68.5520 &  0.31 & 0.40 &   24.6   &  --             &   --              \\
      \hline                                                    
      SReg & S.01 & $-$9.27134 & 2.06101 &  0.08 & 0.11 &   44.8   &  --             &   --              \\
      SReg & S.02 & $-$7.57631 & 6.09103 &  0.04 & 0.04 &   43.6   &  --             &   --              \\
      SReg & S.03 & $-$5.31655 & 7.60502 &  0.14 & 0.23 &   43.3   &  --             &   --              \\
      SReg & S.04 & $-$1.88907 & 1.84101 &  0.16 & 0.31 &   43.2   &  --             &   --              \\
      SReg &S.05 & $-$1.82468 &$-$0.05600 &  2.55 & 3.59 &   42.5   & 0.63~$\pm$~0.13 & $-$65~$\pm$~6 \\
      SReg & S.06 & $-$1.61411 &$-$2.82499 &  1.35 & 1.46 &   41.9   & 1.15~$\pm$~0.24 & +58~$\pm$~6 \\
      SReg & S.07 & +0.00000 & 0.00000 &  5.53 & 6.77 &   41.6   & 0.28~$\pm$~0.04 & $-$32~$\pm$~4 \\ 
      SReg & S.08 & $-$34.4941 &$-$1.73098 &  0.10 & 0.14 &   40.2   &  --             &   --              \\
      SReg & S.09 & $-$29.9659 &$-$3.77899 &  0.04 & 0.05 &   40.0   &  --             &   --              \\
      SReg & S.10 & $-$35.9028 &$-$3.57598 &  0.05 & 0.07 &   39.3   &  --             &   --              \\
      \hline
    \end{tabular}
    \label{maserprop}
>From left to right, the following 
maser parameters are presented: region of location (Reg); feature label 
(Label); right ascension offset, in reference to Figure~\ref{maserpos} 
($\Delta\alpha$); declination offset, in reference to Figure~\ref{maserpos} 
($\Delta\delta$); peak intensity ($I$); integrated intensity ($\int I$); 
radial velocity (V$_{lsr}$); P; EVPA.
  \end{minipage}
\end{table*}

\subsection[]{Linear Polarization}
\label{linpol}

We detected linear polarization for three of the 11 maser features in the SReg 
(S.05, S06 and S07). The degree of polarization and polarization angles of 
each feature are listed, with their respective errors, in columns 8 and 9 of 
Table~\ref{maserprop}. Those values correspond to the results given by the 
brightest channel of each feature. The P error is given by the rms taken from 
the P image, scaled by the intensity peak. The EVPA error was determined 
using the expression $\sigma_{EVPA} = 0.5~ \sigma_{P}/P \times 180^\circ/\pi$ 
\citep{wardle74}. 

In Figure~\ref{maserpos}.d, the black lines show the direction of the 
linear polarization of features S.05, S.06 and S.07. The length of 
the lines are proportional to the linear polarized intensity of each 
feature. 

\begin{figure}
\centering
\begin{tabular}{cc}
  \includegraphics[width = 88 mm]{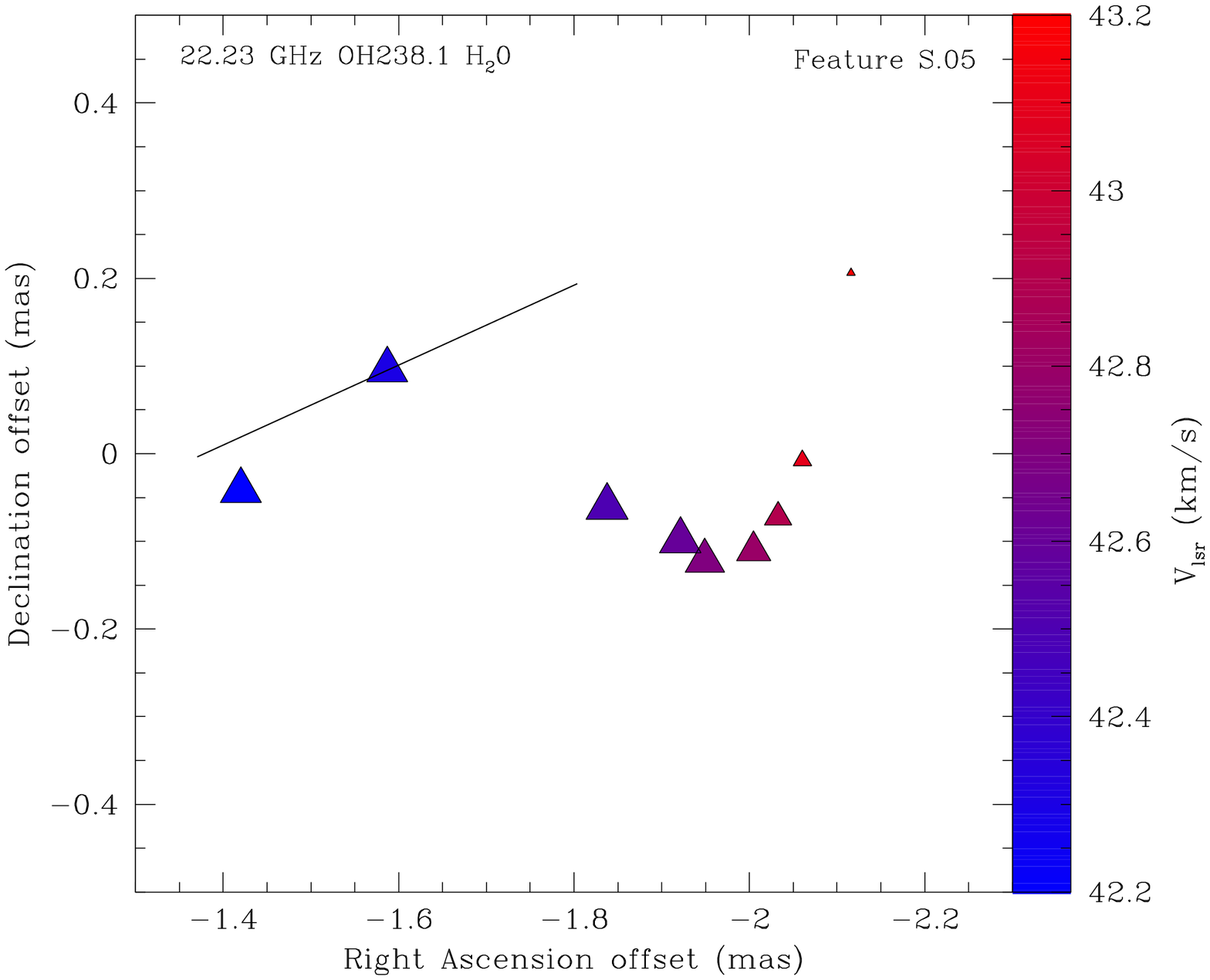} \\
  \includegraphics[width = 88 mm]{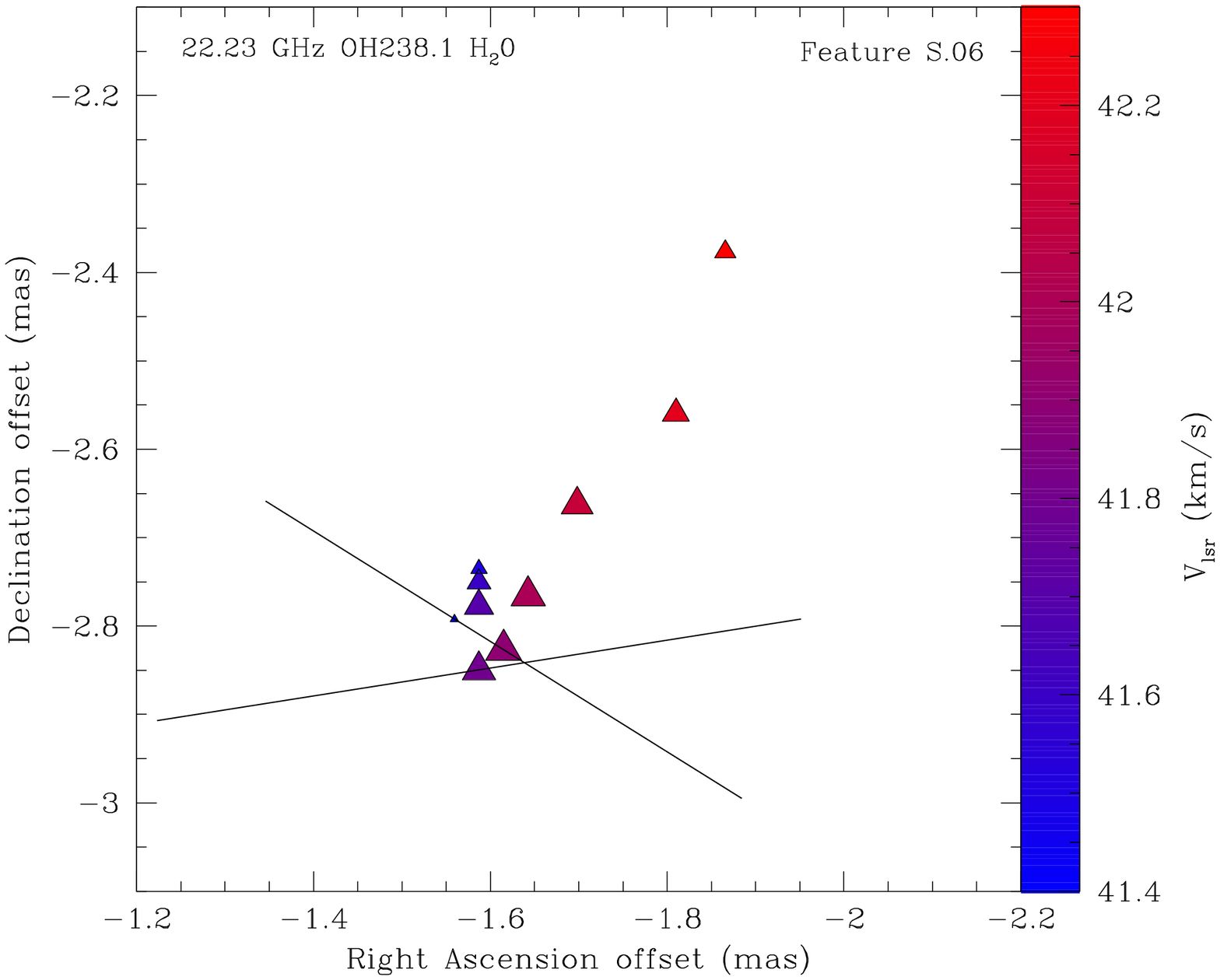} \\
  \includegraphics[width = 88 mm]{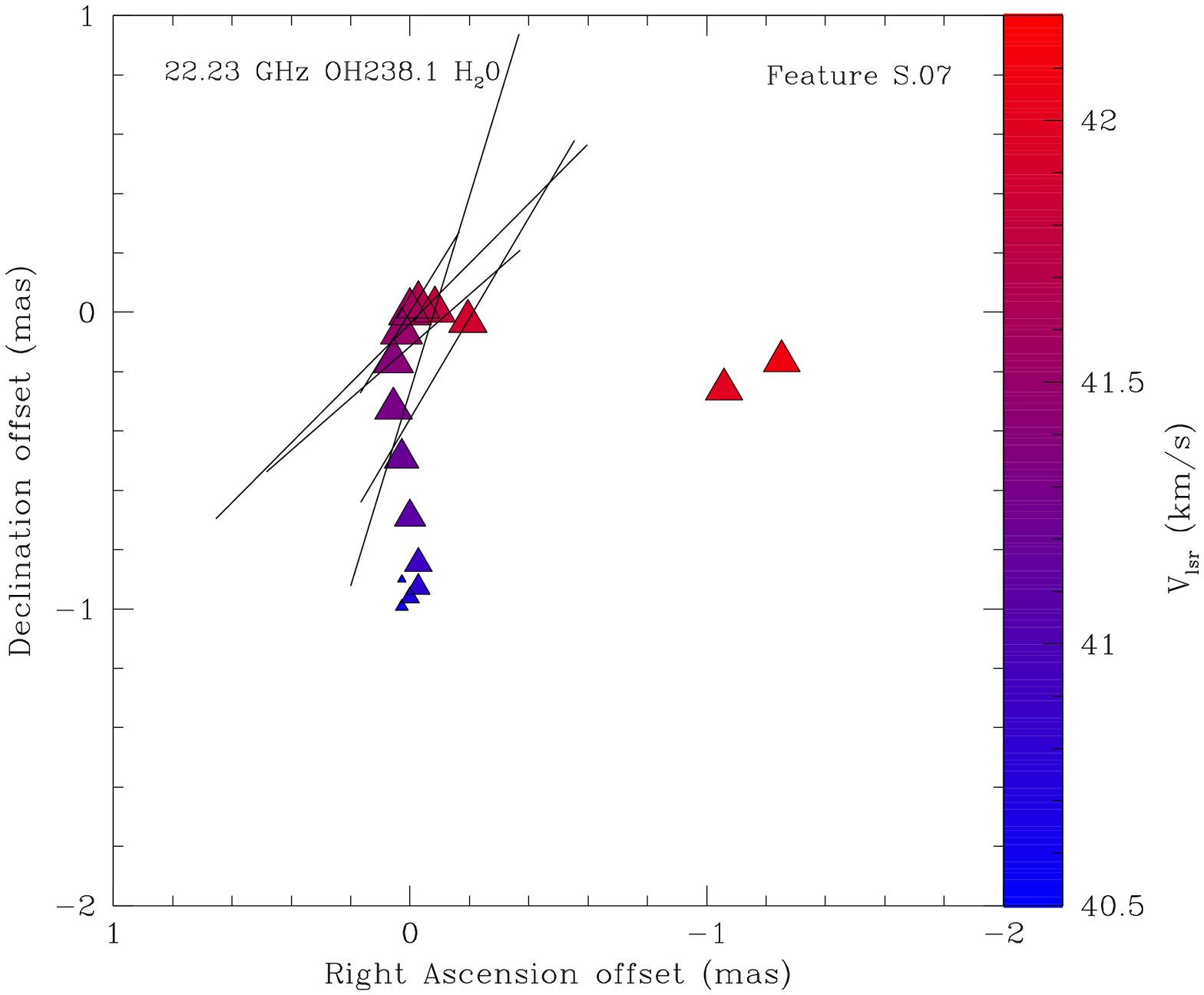} \\ 
\end{tabular}
\caption[]{Maser spots of the features S.05, S.06 and S.07. The triangle 
sizes are scaled to the intensity of each spot emission; the color scale 
shows how the velocity varies at each spot; and the black lines show the 
direction of linear polarization for those spots that have survived to the 
polarization threshold cut. The size of the black lines are scaled to the 
linearly-polarized intensity measured for each spot.}
  \label{linpolchan}
\end{figure}

In Figure~\ref{linpolchan}, for each feature for which linear 
polarization was detected, we show the polarization vectors across all individual 
channels. Again the length of the vectors are scaled in proportion to 
the linearly-polarized intensity. 

\subsection{Circular Polarization}
\label{sectcircpol}

The total intensity ($I$) and circular polarization ($V$) spectra of 
the maser features were used to perform the Zeeman analysis described by 
\cite{vlemmings02}. In this approach, the fraction of circular polarization, 
$P_V$, is given by
\begin{eqnarray}
P_V &=& (V_{max}~-~V_{min})/I_{max} \nonumber \\
    &=& 2~\times A_{F-F'}~\times B_{||[Gauss]} /\Delta v_{L[km s^{-1}]}
\end{eqnarray}
where $V_{max}$ and $V_{min}$ are the maximum and minimum of the model fitted 
to the $V$ spectrum and $I_{max}$ is the peak flux of the emission. $A_{F-F'}$ 
is the Zeeman splitting coefficient, whose exact value depends on the relative 
contribution of each hyperfine component of the \h2o 6$_{1,6}-$5$_{2,3}$ 
rotational maser transition. We adopt the value $A_{F-F'} = 0.018$, which is the typical 
value reported by \citet{vlemmings02}. $B_{||[Gauss]}$ is the projected magnetic 
field strength along the line of sight, and $\Delta v_L$ is the full- width 
half-maximum of the $I$ spectrum. Although the non-LTE analysis in 
\citet[][]{vlemmings02} has shown that the circular polarization spectra 
are not necessarily strictly proportional to $dI/d \nu$, using $A_{F-F'}$ 
determined by a non-LTE fit introduces a fractional error of less than 
$\sim$20$\%$ when using Eq.~1. To increase the signal-to-noise ratio
of the $I$ and $V$ spectra, we smoothed the data by applying a running average over three 
consecutive channels. We note that the same result, albeit at lower signal-to-noise ratio, was obtained using the non-smoothed spectra.

\begin{figure}
\centering
\begin{tabular}{cc}
  \includegraphics[width = 42 mm]{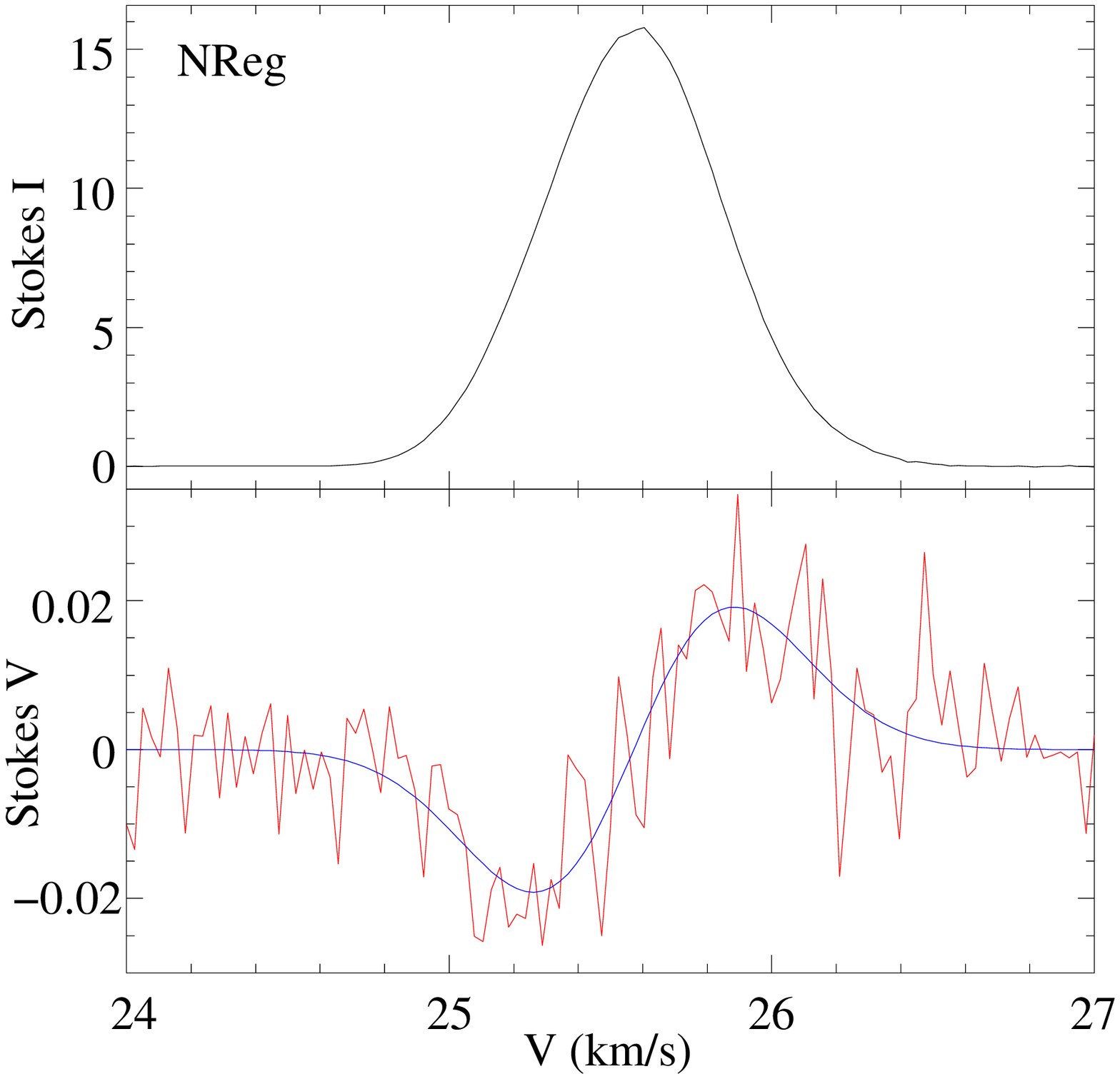} &
  \includegraphics[width = 42 mm]{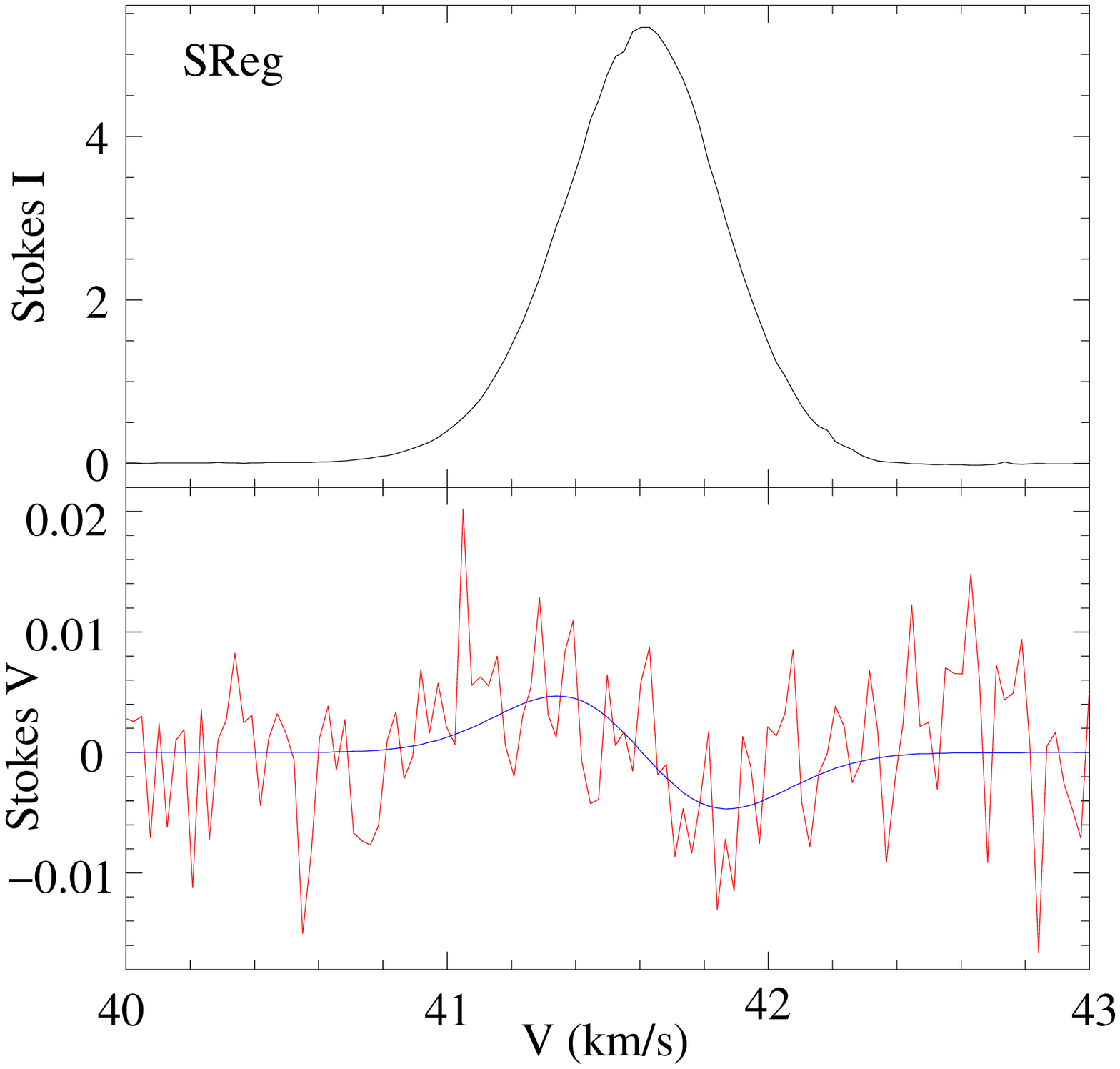} \\
\end{tabular}
\caption[]{Spectra of the Stokes $I$ (top) and $V$ (bottom). On the left 
hand side we present the spectra on the spatial position of the peak emission 
of the NReg. On the right hand side we show the spectra on the spatial 
position of the peak emission of the SReg. From the best fit model (blue line) 
we obtained B$_{||}$(NReg)~=~44~$\pm$~7~mG and B$_{||}$(SReg)~=~$-$~29~$\pm$~21~mG.}
  \label{circpol}
\end{figure}

In Figure~\ref{circpol} we show the spectra from Stokes~$I$ and $V$, with 
$aI$ subtracted. Over-plotted on the $V$ spectrum (red), we show the model 
curve that best fits the spectrum (blue). From this model fit, we derive a magnetic field strength of B$_{||}$(NReg)~=~44~$\pm$~7~mG and 
B$_{||}$(SReg)~=~$-$~29~$\pm$~21~mG. The reported errors are based on the single channel rms using Eq.1, but similar 
errors were found using a chi-square analysis of the model circular 
polarization fitted to the smoothed spectra. We also further confirmed 
the result for both maser features by performing the chi-square analysis 
on spectra smoothed to different spectral resolutions (up to 
$\sim0.2$~km~s$^{-1}$). This leads us to conclude that the detection 
in the SReg is  marginally significant. The flux densities  of the other 
28 features, however, where not sufficient for a detection of circular polarization. 

\section{Discussion} 
\label{discussion}

\subsection{Kinematics} 
 \label{kinematic}

Assuming that the maser features observed in this work, at the epoch of March 1$^{st}$
2009, correspond to approximately the same features detected by D07 on November 
24$^{th}$ 2002, we can estimate the apparent proper motion  between both 
observational epochs. We compared the distance between the mean position of all 
features located in region 2.A to the mean position of all features 
located in region 2.D in our data (Figure~\ref{masercomp}),  against the 
distance between the two corresponding points from D07. We present 
these mean positions and the offsets between the two observational epochs in 
Table~\ref{kinetic}. The offsets\footnote{These offsets occur in tangent 
plane project in the sky. So, strictly, the $\alpha$ and $\delta$ listed in 
Table~\ref{kinetic} correspond to the coordinate angles of these projections.} 
in $\alpha$ and $\delta$ give a total offset of 14.4~mas which, at a distance 
of 1540~pc, corresponds to 22.2~AU. Accordingly, for the 
time elapsed between the observations (2288~days), the velocity of the 
separation between the two mean points on the sky is computed to be 2.3~mas/year, or 
16~km/s. Assuming the inclination of the bipolar flow is 
$i$=36$^\circ$ \citep{kastner92,shure95}, the real separation velocity is 
21~$\pm$~11~km/s. 

Furthermore, we considered the overall distribution of the maser spots 
for all features that we observed. In Figure~1.a we show that the spots in 
the NReg appear to align well with the direction of the nebular jet. 
The same property is found for the spots from the SReg, as can be seen more clearly in Figure~\ref{linpolchan}. In particular, the spots from feature S.07 
appear to be closely aligned with the direction of the jet. However, despite 
being co-linear with the jet, the \h2o ~maser outflow velocity is 
significantly lower than that reported for the jet, v$_{jet}\sim$ 330~km~s$^{-1}$ \citep{sanchez97}. Hence, our interpretation is that the 
masers are being entrained by the jet but are not located in the jet 
itself. As a consequence, the maser spots have a velocity gradient; with 
higher velocity further from the origin. The nature of the \h2o ~masers in 
\source ~is thus very different from that of the water fountain sources where 
the \h2o ~masers lie at the tip of their fast bipolar outflows \cite[e.g][]{imai02}.

It is important to highlight, however, that  a more precise 
kinematics analysis will require more accurate astrometric 
observations.

\begin{table}
  \centering
  \begin{minipage}{75mm}
    \caption{Coordinates of the center (mean) points of the features located 
in regions 2.A and 2.D, from Figure~\ref{masercomp}; and their offsets between 
the observations from D07 and ours.}
    \begin{tabular}{@{}ccccccccc@{}}
      \hline
      Reference        & Region &$\Delta\alpha$ (mas)& $\Delta\delta$ (mas)  \\
      \hline
      Present work     &  NReg  & +2.3 & +29.2 \\
      Present work     &  SReg  & +9.7 & $-$44.0 \\
      D07 &  NReg  & +2.6 & +26.2 \\
      D07 &  SReg  & +2.4 & $-$34.6 \\
      \hline
      Offset           &  NReg  & $-$0.3 & +2.9 \\
      Offset           &  SReg  & +7.4 & $-$9.4 \\
      Offset           &  Total &  7.7 & 12.3 \\
      \hline
    \end{tabular}
    \label{kinetic}
  \end{minipage}
\end{table}

\subsection{Polarization} 
 \label{poldiscut}

We detected linear polarization for three \h2o maser features, all of them 
located in the SReg (S.05, S.06 and S.07). According to maser
  polarization theory, the 
direction of the linear polarization can be either parallel or perpendicular 
to the direction of the magnetic field lines. It is parallel when the angle 
between the magnetic field and the direction of maser propagation ($\theta$) is
less than the van Vleck angle 
($\sim$55$^\circ$), and it is perpendicular when $\theta$ is greater
than this angle \citep{goldreich73}. The percentage of linear polarization is affected by 
$\theta$ and by the degree of saturation. Based on the level of linear 
polarization detected ($\sim1\%$) we cannot conclude definitively in which 
regime - parallel or perpendicular - the masers of \source ~occur. 

The linear polarization angles of the features S.05, S.06 and S.07 display 
a significant scatter  (Fig~\ref{maserpos}.d). The cause 
of this scatter is not known, but we propose three different 
scenarios: (i)~the scatter could be caused by turbulence, or (ii)~in the case of 
a toroidal magnetic field, the masers could be located at its tangent points; both would explain the different local directions of the field, or (iii) the scatter could be caused by internal Faraday rotation \citep{faraday45}. Under the
effect of Faraday rotation, the linear polarization angle of the emission 
($\Phi$) is rotated by
\begin{eqnarray}
\Phi[^\circ]=2.02~\times 10^{-2}~L[AU]~n_e[cm^{-3}]~B_{||}[mG]~\nu^{-2}[GHz]
\end{eqnarray} 
where $L$ is the path-length of the maser emission through a medium
with magnetic field $B_{||}$ and electron density $n_e$ at a frequency
$\nu$. By assuming typical values $L$$\sim$10~AU and
$B_{||}$$\sim$50~mGfor the \h2o masers, Eq.~2 becomes:
$\Phi$$\sim$0.02$^\circ$~$\times$~$n_e$. For the Faraday rotation angle to be 
of the order of the observed scatter, an electron density of order 10$^{3}$ is required. For example, if $n_e$$\sim$2000~cm$^{-3}$, then 
$\Phi$$\sim$40$^\circ$. For an electron density of this order at this
distance to the central star(s), the fractional ionization would be
significantly higher then for a regular evolved star. However, considering
the pre-PNe nature of \source ~this cannot be completely ruled out. Still, we 
consider either scenarios (i) and (ii) more likely than the internal Faraday 
rotation effect. We note, however, that both scenarios (i) and (ii) differ from 
the results reported by \citet{etoka09} for the OH maser region. The OH masers 
indicate a uniform field that seems to be flaring out in the same direction 
of the jet. However, the \h2o and OH masers occur in very different regions. 
While the current work finds the \h2o masers to be located at $\sim$40~AU from the CS, the OH 
masers are located on a torus with a radius of $\sim$2~arcsec \citep[][]{zijlstra01,etoka08}. 
At 1540~pc this correspond to a distance of $\sim$3100~AU. %{\bf Hereupon, it is 
%interesting to highlight the different behavior of the magnetic field in these 
%two regions. While the field has a poloidal well organized structure 
%further out from the central region, it could be either toroidal or less 
%uniform close to the central star. As the results we got from the \h2o 
%masers reveal the evolution from the last few decades, perhaps it could mean 
%that the magnetic field morphology is evolving in time. In this sense, is 
%the original poloidal field going to be swept out by a more powerful wind?}

We detected circular polarization for two maser features. Based on these
detections, we determined the strength of the magnetic field in each
feature to be (B$_{||}$(NReg)~=~44~$\pm$~7~mG, B$_{||}$(SReg)~=~$-$~29~$\pm$~21~mG). 
If, as argued in Sect.~3.3 that the detection 
in the SReg is  marginally significant, it is evident that the magnetic 
field in the SReg has an opposite sign compared to the field in the NReg - the 
positive sign indicates that the direction of the magnetic field is away 
from the observer, while the negative sign corresponds to a direction towards 
the observer. 

A precise determination of the morphology of the magnetic field is
not possible from the current work, as different field morphologies can fit the circular
and linear polarization results obtained here. To reach a final
conclusion as to the morphology of the magnetic field of \source ~more
sensitive polarization information is necessary.  If, however, we
assume a toroidal\footnote{Note that solar-type and
    dipole fields have $r^{-2}$ and $r^{-3}$ dependencies,
    respectively.} magnetic field
(B~$\propto$~$r^{-1}$), it is possible to extrapolate the field strength to the
stellar surface. By taking the separation between the NReg and SReg,
and assuming that the CS is located centrally between the two, each \h2o
maser region is at $\sim$40~AU from the star. Therefore, from the measured  B$_{||}$ for these
regions, the magnetic field
strength on the stellar surface (taken to have a radius of $\sim$ 1~AU) must be
$\sim$2.5~G.This is consistent with the
  result found for W43A \cite[1.5~G;][]{vlemmings06}. This value
  presents a lower limit if the magnetic field vs. radius relation has
a steeper than toroidal dependence.

\section{Conclusions} 
\label{conclusions}

We detected 30 \h2o maser features towards \source. By comparing the current work with the 
prior observations of \cite{desmurs07}, we can conclude that the features are 
moving away from the central star. The average separation velocity 
of the masers is 21~$\pm$~11~km/s. Furthermore, the masers appear to be 
dragged in the direction of the collimated outflow. This likely indicates 
that the masers arise in the turbulent material that is entrained by the 
jet. 

We detected linear polarization for three \h2o maser features. The large 
scatter in the directions of the linear polarization could be caused by 
turbulence, or could be due to the masers being located close to the  
tangent points of a toroidal magnetic field. The possibility of 
Faraday rotation has also been investigated but, unless the electron 
density is exceptionally high ($\gtrsim$~2000~cm$^{-3}$), is 
ruled out.

Based on the detection of circular polarization for two maser features, we 
determined that the strength of the source magnetic field is B$_{||}$(NReg)~=~44~$\pm$~7~mG and B$_{||}$(SReg)~=~$-$~29~$\pm$~21~mG. 
The exact morphology of the field in the \h2o maser region could however not be 
determined. Assuming a toroidal magnetic field (B~$\propto$~$r^{-1}$), 
the extrapolated magnetic field strength on the stellar surface is 
$\sim$~2.5~G.

Our polarization detections, together with the results from \cite{etoka09}, 
make  pPN \source ~the first evolved star that is known to be a binary 
and in which the presence of a magnetic field is confirmed throughout
the circumstellar envelope. A more comprehensive model of the 
magnetic field morphology and its potential evolution based on a comparison 
of the inner and outer wind will require more sensitive observations.

%______________________________________________________________ 
 
\begin{acknowledgements} 
MLL-F would like to thank Dr. G. Surcis, Dr. R. Franco-Hern\'andez and 
the anonymous referee for their useful suggestions, that helped to improved the 
paper. This research was supported by the Deutscher Akademischer Austausch Dienst (DAAD) and 
the Deutsche Forschungsgemeinschaft (DFG; through the Emmy Noether Research 
grant VL 61/3-1).
\end{acknowledgements} 
 
%______________________________________________________________ 

\end{document}